\documentstyle{elsart}
\input epsf
\epsfverbosetrue
\def\ltsima{$\; \buildrel < \over \sim \;$}
\def\lsim{\lower.5ex\hbox{\ltsima}}
\def\gtsima{$\; \buildrel > \over \sim \;$}
\def\gsim{\lower.5ex\hbox{\gtsima}}
\begin{document}
\voffset-.8cm
\begin{frontmatter}

\title{The New Paradigm for Gamma Ray Bursts: \\ 
a Case of Unethical Behaviour?}

\author[]{A. De R\'ujula}

\address{Theory Division, CERN, CH-1211 Geneva 23, Switzerland}

\begin{abstract}
  One might have hoped that the immediacy and completeness of scientific
  information provided through the internet would have made the wrongful
  appropriation of someone else's ideas ---now
  so easy to detect and document--- a sin of the past. Not so.  Here I
  present evidence, which the reader should judge, of such an apparent
  misconduct by
  Sir Martin Rees,  the British Astronomer Royal, and others.
  Unethical behaviour is not unknown in science. My sole intention is to
call attention to the problem by way of example, in an attempt to
contribute to a more ethical atmosphere, which would in my opinion
be beneficial to the field.

\end{abstract}

\end{frontmatter}

{\it Our lives begin to end the day we become silent about things that matter.}
Martin Luther King, Jr.

\section{Motivation}

It is often said that {\it `battles between scientists are so
  fierce... because there is so little at stake'}.  This is probably true of
most scientific disputes, perhaps even of the matter I shall discuss.  However,
this note is less concerned with scientific issues than with questions of
academic integrity: the proper procedure by which novel ideas may be adopted
from the works of others, the appropriate attribution of priorities, and the
example to be set by those in high office --- in this instance, Sir Martin
Rees, the British Astronomer Royal and Master designate of Trinity College,
Cambridge.

The issue at hand is the understanding of Gamma Ray Bursts (GRBs), intense
but transient showers of high-energy photons (gamma rays)
impinging upon the upper atmosphere several times per day.  GRBs were
discovered in the late 60's by the American Vela satellites and were later
shown to be  {\it cosmological,} in the sense that the mighty {\it engines} making
them are distributed roughly uniformly throughout the visible universe.

What can these {\it  engines} be and what is the {\it  mechanism} by which the
gamma rays are
produced? These two puzzles, often considered as  among the greatest
mysteries of astrophysics, have challenged scientists for decades. Several
different interpretations of GRBs  ---often vague and mutually incompatible models
that were provisionally accepted by most workers in the field--- have
succeeded one another  to address these questions.  Until
recently, none of these schemes adequately explained the data.  However, I
believe that recent observations of GRBs and their theoretical interpretation
are now sufficient to provide {\it  a definitive paradigm}.  The once-mysterious
engines responsible for GRBs are nothing more than {\it  supernova
  explosions}, and the mechanism producing the bursts is {\it  inverse Compton
  scattering} (ICS), the process by which high-energy electrons strike
low-energy photons, thereby uplifting their energy to that of gamma rays,
seen as GRBs\footnote{To be precise, these answers apply most convincingly to 
the best studied GRBs, those of {\it long duration} (seconds or minutes, 
as opposed to fractions of a second).}.

The currently best-studied theories of GRBs are the {\it  Fireball} models 
and the {\it  Cannonball} model. The first set of models is often
considered to be {\it the standard model} of GRBs.
The second, of which I am a coauthor, is generally viewed as 
{\it heretical} (to borrow a term used by one of our anonymous referees).
In spite of their similarly-sounding names, these two models
are completely different in their basic hypothesis, in their description
of the data, and in their predictions.

What I shall discuss here is not the validity
of any particular model, but the ways by which some scientists  rewrite
history by de-emphasizing the contributions of others relative to their
own, or by imposing their own modes of thought upon the community.
I shall also pose the question of whether the main concepts underlying 
what is very likely to become the new {\it  GRB paradigm} will be
attributed to their creators. Today, because scientific articles are virtually
instantaneously `posted on the web' in freely accessible and inviolate
electronic Archives\footnote{Most articles cited here can be found in the
{\it Astrophysics} or (in one case) the {\it High-Energy-Physics-Phenomenology}
section of {\tt http://www.arxiv.org/} Their numbers
(year, month, serial arrival number) are quoted in the reference list.
New versions of an article can be added to the Archives, but 
the older versions cannot be erased, even if the paper
is `withdrawn'.}, one might naively
suppose an affirmative answer. But science is a profoundly human endeavour and
what may appear obvious is not always so.

Whether or not a note such as this should be posted in the Archives is
a debatable  question, alternatives to which I have discusssed with
many colleagues. But a ``higher court'' of science does not exist.
Moreover, the Archives now have a much greater impact than journals: it is
mostly the original version of posted papers that people read, 
without checking the journal versions for changes, even when
the journal version is the one they quote. In my
opinion, what is freely posted in the Archives ought to be discussed in the
Archives, since it is the sole responsability of its authors.

In this note I shall focus on the {\it mechanism} producing GRBs (for
which I have no responsibility) rather than the {\it engine} generating them
(for which development I do claim a shared responsibility). This gives me
the benefit of a somewhat wider perspective.

\section{Facts}

By {\it facts} I mean the statements itemized below, in which
all of the quotations are copied from the Archives on the web.

{$\bullet$}
In an article posted on the web on 14 July, 1994, Nir Shaviv and Arnon Dar
\cite{SD}
argued persuasively 
that ICS is  the mechanism underlying  GRB generation. 
I quote from their abstract: 
 {\it
``Striking similarities exist between high energy gamma ray emission from 
active galactic nuclei and GRBs. They suggest that 
GRBs are generated by} {\bf inverse Compton scattering} {\it from highly 
relativistic electrons in transient jets. Such jets may be produced along 
the axis of an accretion disk formed around stellar black holes (BH) or 
neutron stars (NS) in BH-NS and NS-NS mergers and in accretion-induced 
collapse of magnetized white dwarfs or NSs in close binary 
systems"}\footnote{The quoted production mechanisms
are those thought to give rise to ``Type Ia'' 
supernovae. Our current contention
is that these supernovae are responsible for short-duration GRBs \cite{DD03}.}.
 The emphasis is mine.
In this article Shaviv \& Dar predicted what would be a tell-tale signature
of the ICS mechanism: {\it  a very large polarization of the gamma-rays
of a GRB}\footnote{Shaviv \& Dar state that ICS was first
discussed as a mechanism for GRB production by Shemi \cite{Sh}, but
in the context of (spherical) {\it fireballs}, in which there is no expected
polarization. Actually, ICS was discussed earlier by Zdziarski et al.~\cite{Z}.}.

{$\bullet$} It is difficult to measure the degree of polarization of
the gamma-rays of a GRB. Very recently, however, the RHESSI satellite
(whose primary purpose is to study the Sun) succeeded in measuring 
 \cite{Pol} the
polarization of a GRB\footnote{The data analysis has been criticised
\cite{RF} and defended \cite{Pol2}.}.  The result is $\Pi=80\pm 20\%$. That is,
{\it  the gamma ray polarization is virtually complete, as predicted by 
Shaviv \& Dar.} 

{$\bullet$} From the 90's and until most recently, the generally-accepted
paradigm \cite{Mes,Pi1,Pi2,Wa1}
for the mechanism generating GRBs was {\it  synchrotron radiation}:
light emitted by electrons moving in a magnetic field\footnote{Synchrotron
  radiation can produce a large polarization, but only under the most
  unrealistic circumstances. In my opinion, this is what recent articles on
  the subject show \cite{Lev,Wax,Pir}, although it is not what they say.}.
 Nonetheless, at the time (1994), { Shaviv \& Dar discussed,
 both in person and by mail,
  their then novel idea {\it ---ICS by narrow jets ejected in stellar processes}---
  with Martin Rees, who kindly pointed to them a typografical 
  error in a calculation they had originally quoted and used.} 

{$\bullet$} 
In an article posted on the web on August 14th, 2003 \cite{DD03}, 
Dar and I incorporated in
detail the idea of ICS into the Cannonball model of GRBs
\cite{DD00}, which we have developed in collaboration with Shlomo Dado
\cite{DDDoptical,DDDRadio}.
The large polarization it entails is but one of the many successes of the model.
Not surprisingly, in that paper \cite{DD03} we provide ample
reference to the original idea of Shaviv \& Dar. On the 
%subject of the 
GRB-generating mechanism, we say in the introduction: {\it
``The gamma-rays of a GRB may not be produced by synchrotron radiation
and their polarization may not necessarily imply a strong, large-scale,
ordered magnetic field in their source.  In fact, Shaviv and Dar (1995)
suggested that highly relativistic, narrowly collimated jets ...
may produce cosmological GRBs by inverse Compton scattering (ICS)
of stellar light".}

{$\spadesuit$} 
In an article posted on the web on September 1st, 2003 \cite{ReesPol}, 
Rees and his junior
collaborators state: {\it ``In this paper we show that the detection of a 
high level of linear
polarization does not necessarily imply that the gamma-ray photons
are produced by synchrotron... Compton
(Thomson) scattering of photons can, under the appropriate observing
geometry, produce highly polarized radiation''.} These statements are
very similar to those of the previous item.

{$\spadesuit\spadesuit$}
Even clearer is the {attribution} by Rees and collaborators, in the
same article \cite{ReesPol}, of the idea that Compton scattering is
the GRB-generating mechanism. They write: 
{\it ``The possibility that the gamma-ray photons in the prompt phase of GRBs
are due to the bulk Compton up-scatter of UV field photons was} {\bf  first
discussed} {\it by Lazzati et al. (2000; see also Ghisellini et al. 2000)"}. The
emphasis is mine and the quoted papers \cite{Lazz,Ghis}
are also by Rees and collaborators. The basic idea
was not {``first discussed"} by them, but was thoroughly
studied nine years earlier
 by Shaviv \& Dar. This fact cannot have escaped their attention, for 
 they cite our recent paper \cite{DD03}, though in a 
disparaging manner: 
 {\it ``For $\Gamma \gsim 10$ the curves are
indistinguishable and can be approximated with Eq. 1 with
an accuracy of $2\%$ for $\Gamma=10$ and $0.1\%$ for $\Gamma\ge100$
(see also Dar \& De Rujula 2003, who consider only this limiting
case)''}\footnote{The electrons' {\it Lorentz factor}  is
$\Gamma\equiv 1/\sqrt{1-v^2/c^2}$. In ICS, the energy of the struck photon
is uplifted by a factor that can reach a maximum of $2\,\Gamma^2$.
The Eq. 1 of \cite{ReesPol} is Eq. 28 of Shaviv \& Dar, written in 
its more precise
form. It also appears in this form as Eq. 19 in \cite{DD03}, 
and in many standard books on quantum
electrodynamics.}. Rees and collaborators
neglect to inform the reader that $\Gamma\ge100$ is the {\bf  only}
relevant case, even for their analysis\footnote{The deviations from
the approximate result are of order $1/\Gamma^2$: it is not necessary
to draw curves to convince oneself that, even for $\Gamma=10$,
the precision of the approximation is more than sufficient.
Moreover, for low $\Gamma$, other uncertainties are crucial, e.g.~the
directional distribution of target photons. Thus, in my opinion,
these calculations in \cite{ReesPol} are not only unnecessary,
but also misleading.}.

{$\spadesuit\spadesuit\spadesuit$}
In their paper of the year 2000 \cite{Lazz}, Rees and collaborators state:
{\it The Compton drag effect has already been invoked for GRBs by
Zdziarski et al.~(1991) \cite{Z} and Shemi (1994) \cite{Sh}... none of
these scenarii was able to account for all the main properties of GRBs.}
They do not quote the work of Shaviv \& Dar (1995), which they would 
have no good reason to dismiss. By September 2003, as we have seen, 
Rees and collaborators refer to the Compton mechanism as something
that they `first discussed'. One of the very many reasons \cite{DD03}
 making this mechanism so 
compelling in comparison to any others
is the observation of a high polarization, a result 
announced on the web in May 2003 \cite{Pol}.

$\bullet$
Is the behavioural pattern described above unusual in
the field? No. In the concluding section of Dar \& De R\'ujula \cite{DD03}, 
after itemizing many similar grievances, we dare to state:
{\it ``The Cannonball model} [ours] {\it may of course be wrong, but it is 
successful.  It is not at all inconceivable that the Fireball models}
[theirs]
{\it continue to incorporate and ``standardize'' other aspects of the Cannonball 
model".} In Section 19.1 we referred more specifically to the earlier work of Rees
et al.~\cite{Lazz} in which they `first discuss' the ICS mechanism: 
{\it ``The authors noticed that, for
a large $\Gamma$, the characteristic GRB energies could be explained
(Shaviv \& Dar 1995); but ---perhaps because at the time fireball
ejecta were still considered to be broad and to point at the observer---
they overlooked the crucial prediction of Shaviv and Dar: a large polarization".}
{This time Rees and collaborators \cite{ReesPol}
no longer missed that crucial prediction} \cite{SD}.

\section{Later and earlier developments}

On September 3rd, 2003, Dar sent an e-mail to Rees,
the senior author of the above {appropriation} of ideas, expressing
his surprise and stating: {\it ``With due respect, Sir Martin Rees, I think that 
your position as the Royal
Astronomer of Britain requires setting an example of proper scientific
conduct and more responsibility . I expect you to withdraw your paper and
I expect an apology."} Rees answered
promptly, saying that {\it ``when we revise our paper we  will add a 
reference to your 1995 paper and  change the wording of our  
comment on your  2003  preprint''.}

The response of Rees is inadequate. There is a clear difference
between inadvertently forgetting a relevant reference and knowingly
attributing someone else's ideas to oneself, while paraphrasing the
original authors' principal hypothesis, conclusions and results.
{No one}, least of all the Astronomer Royal, should feel
licensed to write a paper \cite{ReesPol} in which (at least in the
original version posted in the web) the ideas of
others are simply {appropriated}. Even if this 
incorrect attribution of authorship were
a momentary lapse, any responsible researcher,
having written such a paper, should be
apologetic and glad to withdraw it or ---at the very least--- to
correct its posted version {\it inmediately}. The response of 
Rees to Dar is not the solution but a further symptom of the problem.
Thus, on September 4th, 2003, I sent Rees a strongly-worded e-mail asking 
him for an apology (which Dar did not obtain) and again requesting that 
the article  be withdrawn. A copy of the message was sent to Rees
by registered mail a few days later.  I have had no response.
As I wrote to Rees, I am left with no choice but to make this issue public.

Fearing a serious clash between astronomers (Rees) and particle physicists
(me), a common friend of ours wrote to Rees expressing his concerns.
Rees, in his very kind answer (which I was explicitly allowed by Rees to see), 
writes: {\it ``It's very likely that  the  paper will need revising on the basis of 
referee's comments. Indeed it may of course be  rejected -- and
undoubtedly will be if Dar and de Rujula's assessment is well-based".}
I disagree with this transfer of responsibility to an anonymous referee.
It is the duty of the authors and of the editor, who in this case had
been informed of the issue. By October 2nd the paper \cite{ReesPol}
had been accepted, with the proviso that proper references be given.
That will not affect the subsequent
attribution of the original ideas by other workers in the field,
particularly since, to date, the original version of \cite{ReesPol}
is still the only one posted on the web. 

Let me give three other pertinent instances of inadequate citation. 
When the gamma-ray signal of a GRB is over, the event  is not.  The GRB
is followed by an {\it afterglow} (AG).   That is, the source continues to
shine photons of lower energy,
and remains  observable for a very long time. GRB afterglows have been
observed at many frequencies, from X-rays to visible light, to radio.  As far
as I know, the first prediction  of AGs is  due to Jonathan
Katz \cite{Katz1,Katz2,Katz3}, who wrote: {\it ``The debris' energy degrades. Its
  radiation shifts to lower frequencies and increases in duration, and may be
  observable at frequencies from X-rays down to radio.''}\footnote{Paczynski
  \& Rhoads \cite{PR} independently predicted a {\it radio} AG, analogous to
  that of SN remnants.}  The predicted AGs were discovered in 1997 \cite{AG}
and have played a crucial role in the understanding of GRBs.  However, the
prediction of AGs is systematically attributed to M\'esz\'aros and Rees
\cite{MR}, who, some four years after Katz, wrote: {\it ``We find that the
  resulting cosmological GRB remnants ... produce significant fluxes of
  softer radiation, mostly X-rays and optical, but in some cases radio as
  well''.} These authors do cite Katz \cite{Katz1,Katz3}, not for  
the prediction of AGs (which is
what M\'esz\'aros and Rees allege to make), but in 
%an unrelated ``bus reference''\footnote{A bus reference is one in which as 
%many authors or collaborations are quoted as it is possible to fit in a bus.}.
a collective reference unrelated to AGs.
In this case, peer review did not ensure that the earlier work of
Katz was properly aknowledged.

The { self-attribution} by Rees and collaborators
of ICS as the GRB-generating mechanism and of the subsequent polarization
{has born fruit}, for they are already quoted in the literature for
both items. For instance,
Sikora et al.~\cite{Sik}, state: {\it ``Proposed scenarios include...
and} {\bf Compton scattering}{\it ... (Lazzati et al.~2000 \cite{Lazz} and references 
therein}).
{\it If the polarization measured by the RHESSI satellite is real... the Compton
mechanism can still work (... Lazzati et al.~2003 \cite{ReesPol}), provided 
that the jet has an opening angle $<1/\Gamma$."}

Incidentally, the last-cited condition is a complete novelty in Fireball models,
while it is satisfied by construction in the [uncited]
Cannonball model, in which the only relevant angle is
the observer's viewing angle. To
accommodate the data, the original spherical ``Fireball" models had to evolve into 
 ``Firecone'' models \cite{cones,cones2}. Firecones were to emit gamma rays as
idealized torches, that is, within a  cone with a given
``jet'' opening angle. 
 For years, these models were  {\it anthropoaxial}:
the  observer was placed exactly {\it on the axis} of the cone. In 
Dado et al.~2002 \cite{DDDoptical}, we criticized this improbable
hypothesis. Subsequently, in
%\footnote{Technically speaking, for a uniform distribution
%within the cone, $dN_\gamma/d\cos\theta\propto\Theta[\theta-\theta(cone)]$ and
%$dN_\gamma/d\theta\propto\sin\theta\,\Theta[\theta-\theta(cone)]$. 
%The probability of being on axis vanishes, and is maximal at $\theta=\theta[cone]$.
%In more recent   firecone avatars, observers are being moved to that position 
%\cite{Reesangle,angle,Wax,Pir}.}.
a handful of papers, 
one of them by Rees and collaborators \cite{Reesangle},
the relevance of the viewing angle was at last recognized by the GRB
community. We were not cited in any of those papers for this trivial
but crucial point\footnote{One of
these papers was coauthored by the (not anonymous) referee of our papers 
\cite{DDDoptical,DDDRadio}, who had ample time to paraphrase 
and publish our simple geometrical considerations
while delaying our publications for more than a year.}.
 
 \section{In lieu of a conclusion}

I have felt obliged to describe several instances of what appears to me to be
unethical behaviour on the part of Rees and others.
Whether or not this behaviour is {\it plagiarism} may be a matter of opinion.
Alas, this kind of behaviour is not exceptional in the field, as we
have already documented in Dar \& De R\'ujula \cite{DD03},
particularly in the conclusions and in an appendix on the history 
of the contention that supernovae are the {\it  engines} of GRBs. 

\section{Aftermath}

The first version (v1) of this paper was posted on October 27th. 
Rees and collaborators posted version v2 of their paper in October 29th,
with the comment {\it ``this manuscript is posted following the comments and allegations of Dr. De Rujula.''}. In this v2, full reference is given to Shaviv \&
Dar \cite{SD}
and the comments that I considered inappropriate have been erased.
In this sense, I consider their v2 satisfactory.

The authors of \cite{ReesPol} now refer to \cite{SD} and \cite{DD03} 
as {\it ``the point-source limit.''} Cannonballs \cite{DD00,DD03}
are supposed to initially
contain a dense enclosed radiation field that drives their expansion
during the GRB phase at a radial velocity (in their rest system) 
comparable to (or somewhat smaller than)
the speed of sound in a relativistic plasma ($c/\sqrt{3}$). Since they
travel at $v\approx c$, their ``jet opening angle'', as viewed from the
SN's center (in the SN rest system) is $\theta_j\sim 1/(\sqrt{3}\,\Gamma)$,
or smaller: CBs are ``small", but not pointlike. True enough, we have made the
(good) approximation of neglecting the opening angle relative to the
viewing angle, but cannonballs are not {\it point sources}.

A more significant question of nomenclature ---or tactics--- is the use of the
generic
trademark ``fireball'' to describe contradictory models. The opening
angle of the ``fireballs'' that Rees et al.~choose to study varies from
$\theta_j=0.2/\Gamma$ to $5/\Gamma$ \cite{ReesPol}. 
While the larger of these ``fireballs'' 
are not what we would call cannonballs, they subtend
extremely narrow angles. These ``fireballs''
are very much more similar to cannonballs than to
the original fireballs: concentrical spherical shells spanning
a full $4\pi$ solid angle, or a good fraction thereof,
see \cite{Mes,Pi1,Pi2,Wa1} and references therein. 
In this sense, v2 of Rees et al.~is still a good 
example of a possibility contemplated in \cite{DD03}: 
{\it ``that the Fireball models continue to incorporate and ``standardize'' 
other aspects of the Cannonball model''}. If the opening angle is
further restricted to $\theta_j<1/\Gamma$, as in the statement by
Sikora et al.~\cite{Sik} quoted in section 3, this good example becomes
a perfect example, and we may have been right in having said:
{\it ``fireballs may turn out to 
have always been cannonballs''} \cite{DD03}.

{}


\begin{thebibliography}{9}

\bibitem{SD}
Shaviv, N. J. \& Dar, A. 1995, ApJ, 447, 863; astro-ph/9407039
\bibitem{DD03}
Dar, A. \& De R\'ujula, A. 2003; astro-ph/0308248, submitted to MNRAS
\bibitem{Sh}
Shemi, A. 1994, MNRAS, 269, 1112; astro-ph/9404047
\bibitem{Z}
Zdziarski, A. A., Svensson, R. \& Paczynski, B., 1991, ApJ, 366, 343
\bibitem{Pol}
Coburn, W. \& Boggs, S. E., 2003, Nature 423, 415; astro-ph/0305377
\bibitem{RF}
Rutledge, R. E. \& Fox, D. B., 2003; astro-ph/0310385
\bibitem{Pol2}
Boggs, S. E. \& Coburn, W., 2003; astro-ph/0310515
\bibitem{Mes}
M\'{e}sz\'{a}ros, P. 2002, ARA\&A, 40, 137; astro-ph/0111170
\bibitem{Pi1}
Piran,  T.  1999, Phys. Rep., 314, 575;  astro-ph/9907392
\bibitem{Pi2}
Piran,  T.  2000, Phys. Rep., 333, 529; astro-ph/9907392
\bibitem{Wa1}
Waxman, E. 2003, In {\it Supernovae and Gamma Ray Bursters}, 
ed. K. W. Weiler, Lecture Notes in Physics, Springer-Verlag (in press);
astro-ph/0303517
\bibitem{Lev}
Eichler, E., \& Levinson, A., to appear in ApJL; astro-ph/0306360
\bibitem{Wax}
Waxman, E. 2003, Nature 423, 388; astro-ph/0305414
\bibitem{Pir}
Nakar, E., Piran, T. \& Waxman, E. 2003; astro-ph/0307290
\bibitem{DD00}
Dar, A. \& De R\'ujula, A., 2000, rejected for publication in A\&A; astro-ph/0008474
\bibitem{DDDoptical}
Dado, S., Dar, A. \& De R\'ujula, A. 2002, A\&A, 388, 1079; astro-ph/0107367
\bibitem{DDDRadio}
Dado, S., Dar, A. \& De R\'ujula, A. 2003, A\&A, 401, 243; astro-ph/0204474
\bibitem{ReesPol}
Lazzati, D., Rossi, E., Ghisellini, G. \& Rees, M.; astro-ph/0309038,
accepted for publication in MNRAS
\bibitem{Lazz}
Lazzati, D., Ghisellini, G., Celotti, A. \& Rees, M. 2000, ApJ, 529, L17; 
astro-ph/9910191
\bibitem{Ghis}
 Ghisellini, G., Lazzati, D., Celotti, A. \& Rees, M. 2000, MNRAS, 316, L45;
 astro-ph/0002094
 %Woosley, S. E. 1993, BAAS, 25, 894
\bibitem{Katz1}
Katz, J. I., 1994; ApJ, 422, 248; astro-ph/9212006 
\bibitem{Katz2}
Katz, J. I.; astro-ph/9311015
\bibitem{Katz3}
Katz, J. I., 1994, ApJ, 432, L107; astro-ph/9312034
\bibitem{PR}
Paczynski, B. \& Rhoads, J. E., 1993, ApJ, 418, L5; astro-ph/9307024
\bibitem{AG}
Costa E., et al., 1997, Nature,  387, 783; astro-ph/9706065
\bibitem{MR}
M\'esz\'aros, P. \&  Rees, M. J., 1997, ApJ, 472, 232; astro-ph/9606043
\bibitem{Sik}
 Sikora, M. et al.;~astro-ph/0309504
\bibitem{cones}
Rhoads, J. E. 1999, ApJ, 525, 737; astro-ph/9903399
\bibitem{cones2}
Frail, D. A., et al. 2001, ApJ, 562, L55; astro-ph/0102282
\bibitem{Reesangle}
Rossi, E., Lazzati, D. \& Rees, M. J. 2002, MNRAS, 332, 945; 
astro-ph/0112083
%\bibitem{Venice} {\it ``GRBs and the sociology of science"},
%De R\'ujula, A. 2003, in {\it Tenth International Workshop on Neutrino
%Telescopes}, Venice. Edited by Milla Baldo Ceolin. Published by
%the Istituto Veneto di Scienze, Lettere ed Arti, edizioni Papergraf.
%Vol. 2, p. 539; hep-ph/0306140
%\bibitem{DPLB}
%Dado, S.,  Dar, A. \& De R\'ujula, A. 2003, PLB, 562, 121; astro-ph/0211596


\end{thebibliography}
\end{document}